# Superoutbursts of the SU UMa-type dwarf nova CP Draconis

Jeremy Shears, David Boyd, Denis Buczynski, Brian Martin, David Messier, Ian Miller, Arto Oksanen, Jochen Pietz, David Skillman, Bart Staels and Tonny Vanmunster

## Abstract

Analysis of observations of the SU UMa-type dwarf nova, CP Dra, between February 2001 and April 2009 has revealed 15 outbursts, at least eight of which were superoutbursts. The supercycle length is 230+/-56 d. We report photometry of the 2001 and 2009 superoutbursts which shows that they were remarkably similar to each other in terms of the profile of the outburst light curve and the evolution of the superhumps. The outburst amplitude was 5.5 magnitudes and the $P_{sh}$ during the plateau phase was measured at 0.08335(31) and 0.08343(21) d, respectively. In both cases, $P_{sh}$ decreased during the course of the outburst and there is some evidence that there was an abrupt change corresponding to the point in the plateau where a slow fade begins. The 2001 superoutburst was caught during the rise to maximum and during this period we found that the $P_{sh}$ was significantly longer than during the subsequent stages. Superhumps of 0.3 magnitude peak-to-peak amplitude were present at the height of each outburst and these gradually diminished during the outburst. We also report photometry from the 2002 superoutburst, although coverage was not so extensive, which revealed $P_{sh} = 0.08348(32)$ d consistent with the other two superoutbursts. Analysis of time resolved photometry from the 2009 outburst revealed evidence for an orbital period of 0.08084(86) d, giving a fractional superhump period excess $\varepsilon = 0.032(3)$ d.

## Introduction

Dwarf novae are a class of cataclysmic variable star in which a white dwarf primary accretes material from a secondary star via Roche lobe overflow. The secondary is usually a late-type main-sequence star. In the absence of a significant white dwarf magnetic field, material from the secondary passes through an accretion disc before settling on the surface of the white dwarf. As material builds up in the disc, a thermal instability is triggered that drives the disc into a hotter, brighter state causing an outburst in which the star apparently brightens by several magnitudes [1]. Dwarf novae of the SU UMa family occasionally exhibit superoutbursts which last several times longer than normal outbursts and may be up to a magnitude brighter. During a superoutburst the light curve of a SU UMa system is characterised by superhumps. These are modulations in the light curve which are a few percent longer than the orbital period. They are thought to arise from the interaction of the secondary star orbit with a slowly precessing eccentric accretion disc. The eccentricity of the disc arises because a 3:1 resonance occurs between the secondary star orbit and the motion of matter in the outer accretion disc. For a more detailed review of SU UMa dwarf novae and superhumps, the reader is directed to references 1 and 2.

In this paper we present photometry of the 2001, 2002 and 2009 superoutbursts of the SU UMa system, CP Dra, from which we measure the superhump and orbital periods. From an examination of the light curve between these superoutbursts, we investigate the length of the supercycle, i.e. the time between superoutbursts

**History of CP Dra**

CP Dra was first reported in a note to Astronomische Nachrichten written by Mrs Dorothea Roberts in 1913, following an examination of a plate taken at her late husband's observatory at Crowborough, UK, on 1904 April 6 [3]. Isaac Roberts (1829 – 1904) was an original member of the BAA and is well known as a pioneer of astrophotography, much of which he conducted at his Starfield observatory at Crowborough. Both Isaac and Dorothea shared the work of analysing and measuring the vast collection of plates taken over many years, with Dorothea continuing this task after his death. An excellent biography of Isaac Roberts in an earlier JBAA provides an insight into Isaac and Dorothea's achievements [4]. Dorothea noted the star was close to the galaxy NGC 3147, but was absent from a plate of the same region exposed in 1895 as well as from the Greenwich Astrophotographic Chart. She went on to say "*The star appears to be of the 13-14 magnitude, it seems real. Is it a variable star? Is it a nova? May I appeal to powerful telescopes for enlightenment?*"

There has been some speculation that Dorothea's mystery object was a supernova and in the 59th list of Variable Stars [5], CP Dra is listed a "supernova or nova near NGC 3147", with a reference to her observation as well as to a possible supernova discovery by Altizer in 1972. Altizer noted that the object "of stellar appearance" was $m_{pv}$ = 15.5 at discovery on 1972 Jan 10.365 [6]. An observation by J.R Dunlap showed that it had faded to mag 16 on Jan 12 [7]. Notwithstanding the possible misidentification of CP Dra as a supernova, NGC 3147 has produced four confirmed supernovae: SN 2008fv, 2006gi, 1997bq (discovered by BAA member Stephen Laurie) and 1972H [8]. Given this apparent relatively high rate of supernova production, observers are encouraged to be vigilant for further supernovae in this galaxy when they observe CP Dra.

CP Dra was finally confirmed to be a dwarf nova by Kolotovkina [9] who examined archival plates and found 6 outbursts during the 1970s (Table 1). The mean outburst interval was 355 days, with a range of 296 to 478 days, whilst the outburst duration was typically 10 to 12 days. The maximum brightness was $m_{pg}$ = 15.10 compared to a quiescence value of about magnitude 20 which Kolotovkina estimated by examination of Palomar Observatory plates. The USNO-B1.0 catalogue has magnitude B=20.5 and R=19.9 [10]. Based on these catalogued magnitudes we take the quiescent magnitude to be ~20.

Superhumps were detected for the first time by Tonny Vanmunster during the 2001 outburst, identifying CP Dra as a dwarf nova of the SU UMa family [11]. Subsequent to this outburst, CP Dra has been intensively monitored. The object was added to the BAA Variable Star Section's Recurrent Objects Programme in 2001 [12]. The aim of this Programme is to encourage observations of poorly studied eruptive stars where outbursts occur at periods of greater than a year (it was removed from the Programme in July 2006 after a multiplicity of outbursts had been detected). An image of CP Dra in outburst is shown in Figure 1.

**The length of the supercycle**

The AAVSO International Database contains over 5000 observations of CP Dra, the vast majority being made during and after the 2001 outburst. We have plotted the observations made between 2001 Feb 23 and 2009 April 15 as Figure 2, where negative ("fainter than") observations have been omitted for clarity. We also examined images of SN2008 in NGC 3147 posted on the Bright Supernovae web pages and found CP Dra in outburst on 2008

Oct 27 in CCD images taken by Jean Marie Llapasset of Perpignan, France [13] and his photometry of this outburst was included.

In order to identify possible outbursts in the data, we noted when the star was brighter than 16.0. Table 2 shows that 15 such outbursts have been recorded, although it is likely that others have been missed due to incomplete coverage. Those outbursts which were longer than 7 days, or which were confirmed to exhibit superhumps, were assumed to be superoutbursts. We note that several outbursts only received 1 or 2 observations and it is possible that some of these were also superoutbursts.

The intervals, $\Delta T$, between each of the eight superoutbursts are shown in Table 3. The mean outburst interval is 421 d and the median 405 d. A preliminary superoutburst cycle number was assigned to each observed superoutburst and a linear ephemeris was calculated from the superoutburst times:

$$JD_{max} = 2451978(28) + 230(28) \times E \qquad \text{Equation 1}$$

This suggests a superoutburst period of 230 +/- 28 d. The corresponding O-C diagram is shown in Figure 3. The O-C residuals range over about +/-78 d, or about 34% of the proposed superoutburst period. The standard deviation of the residuals is 56 d and this value probably gives a more realistic idea of when a superoutburst might be seen, rather than the formal error on the superoutburst period [18]. Thus the adopted superoutburst period is 230 +/-56 d. However, the size of the standard deviation, at nearly 2 months, means that it has little predictive value for future superoutbursts. We also note that the curved shape of O-C diagram is suggestive of a decreasing supercycle length, although more data would be required to confirm this.

The only other published estimate of the supercycle length of CP Dra is 400 d in a 2003 paper by Kato et al. [19]. This is presumably based on the interval between the March 2001 and February 2002 superoutbursts, which according to the data in Table 2 were separated by 397 d. Based on our proposed supercycle of 230 d, is it possible that a superoutburst was missed in the intervening period.

**Time resolved photometry**

Unfiltered (Clear, "C") time resolved photometry was conducted during the 2001, 2002 and 2009 superoutbursts using the instrumentation shown in Table 4. Each observer calibrated their own images by dark-subtraction and flat-fielding and then carried out differential photometry. In 2001 they used a range of comparison stars from different sources, in 2002 the comparison star was the one labeled 136 (V=13.575) on AAVSO sequence 1322ebo and in the 2009 outburst observers used comparisons from the same AAVSO sequence. This, coupled with the different spectral sensitivity of the CCD cameras, means that small systematic differences between their results are likely. However, given that the aim of the study was to investigate periodic variations in the light curve, we consider this not to be a significant disadvantage.

**The 2001 superoutburst**

The 2001 superoutburst was first reported by Chris Jones on 2001 Feb 23 [20]. An extensive campaign of photometry was conducted by members of the Center for Backyard Astrophysics (CBA), a worldwide network of small telescopes, and a log of observations is given in Table 5. The complete outburst light curve is shown in the top panel of Figure 4 and its profile is typical of a dwarf nova outburst. The first 8 days correspond to the outburst plateau, following which there was a rapid decline. The outburst lasted at least 10 days (the return to quiescence was not observed, so the duration is not well constrained) and reached magnitude 14.5 at maximum, an outburst amplitude of about 5.5 magnitudes.

We also plot expanded views of the plateau phase in Figure 5 which clearly show the presence of superhumps. We note that during the first night of photometry (see top panel of Figure 5), the star was brightening, indicating the super outburst was just beginning, and the superhumps were growing in amplitude reaching a peak-to-peak amplitude of 0.3 magnitude. Towards the end of the plateau they were 0.2 magnitude amplitude and subsequently 0.1 magnitude during the decline. We note that on the night that the outburst was discovered, CP Dra was actually slightly brighter (14.3 to 14.5) than the following night (~14.8) when the first time series photometry was initiated. This may indicate that there was a precursor outburst which began to fade before triggering the superoutburst, but in the absence of more complete data in the intervening period this must remain speculation.

To study the superhump behaviour, we first extracted the times of each sufficiently well-defined superhump maximum by fitting a quadratic function to the individual light curves. Times of 27 superhump maxima were found and are listed in Table 6. Unfortunately it was not possible to measure maxima during the rapid decline. Having excluded the data from JD 2451965 to 2451966 at the beginning of the outburst when it appeared the period was changing rapidly, we assigned preliminary superhump cycle numbers to these maxima. An analysis of the times of maximum allowed us to obtain the following linear superhump maximum ephemeris:

$$HJD_{max} = 2451966.7203(63) + 0.08335(31) \times E \qquad \text{Equation 2}$$

The observed minus calculated (O–C) residuals for *all* the superhump maxima relative to the ephemeris are shown in the bottom panel of Figure 4. The plot clearly shows that the superhump period was decreasing rapidly during the first part of the outburst between JD 2451965 to 2451966 when the star was still brightening (corresponding to the data which were excluded from the linear analysis which led to Equation 2). A linear analysis of the first 4 superhump times, during this brightening phase, yielded $P_{sh} \sim 0.0885$ d, which is about 6% longer than in the later stages. The data are consistent with a continuous change in period with $dP_{sh}/dt = -9.7(11) \times 10^{-3}$ between JD 2451965 and 2451967. By contrast the period change during the rest of the plateau was an order of magnitude less at $dP_{sh}/dt = -9.2(9) \times 10^{-4}$

To confirm our measurements of $P_{sh}$, we carried out a period analysis of all the data, except those from JD 2451965 - 2451966 when the superhump period was still evolving, using the Lomb-Scargle algorithm in the Peranso software package [21], having subtracted the linear trend from the data. This gave the power spectrum in Figure 6, which has its highest peak at a period of 0.08345(121) d and which we interpret as $P_{sh}$. This value is consistent with our earlier measurement from the times of superhump maxima. The superhump period error estimate is derived using the Schwarzenberg-Czerny method [22]. Several other statistical algorithms in Peranso gave the same value of $P_{sh}$. A phase diagram of the data from the plateau phase, folded on $P_{sh}$, is shown in Figure 7. This exhibits the typical profile of superhumps in which the rise to superhump maximum is faster that the decline.

While the two methods of measuring $P_{sh}$ gave consistent results, it is not unusual for the time of maxima analysis to result in a more accurate method of tracking periodic waves in dwarf novae as it is less troubled by changes in amplitude than period analysis techniques. Hence we adopt our value of $P_{sh}$ from the time of maxima analysis as $P_{sh}$ = 0.08335(31) d.

**The 2002 superoutburst**

The 2002 outburst received much less attention that the one in 2001, resulting in only partial coverage of the outburst light curve corresponding to the plateau phase (Figure 8, top panel). A log of time resolved photometry observations is shown in Table 7. Superhumps with were observed, initially with a peak-to-peak amplitude of 0.3 magnitudes, gradually diminishing during the plateau phase (Figure 9). As before, we measured the times of 7 superhump maxima (Table 8) and obtained the following linear ephemeris:

$$HJD_{max} = 2452361.3738(52) + 0.08348(32) \times E \qquad \text{Equation 3}$$

The observed minus calculated (O–C) residuals for the superhump maxima relative to this ephemeris are shown in the bottom panel of Figure 8. The plot clearly shows that the superhump period was decreasing, corresponding to a continuous change in period with $dP_{sh}/dt = -1.35(95) \times 10^{-3}$

Using the Lomb-Scargle algorithm in Peranso we measured $P_{sh}$ = 0.08333(98) d (data not shown), which is consistent with the linear analysis of the superhump times. Again we adopt the superhump period from the latter as $P_{sh}$ = 0.08348(32) d.

**The 2009 superoutburst**

The complete outburst light curve, shown in the top panel of Figure 10, is very similar to that of the 2001 outburst, with an amplitude of about 5.5 magnitudes. 14 days after the detection of the outburst the star was at 19.1C, still above the quiescence brightness.

Superhumps were observed throughout the outburst (Figure 11). Initially they had a peak-to-peak amplitude of 0.3 magnitudes, but this decreased during the outburst until they were of about 0.1 magnitude amplitude during the decline. We measured the times of 19 superhump maxima (Table 10) and, as before, fitted these times to a linear ephemeris:

$$HJD_{max} = 2454915.4421(41) + 0.08343(21) \times E \qquad \text{Equation 4}$$

The measured value of $P_{sh}$ = 0.08343(21) d is consistent with that of the 2001 superoutburst (0.08335(31) d).

The O–C residuals relative to this ephemeris are plotted in the bottom panel of Figure 10. The data are consistent with a continuous change in period with $dP_{sh}/dt = -4.1(9) \times 10^{-4}$.

A Lomb-Scargle analysis of the data has its strongest signal at 0.08348(97) d, plus its 1 and 2 c/d aliases, which is also consistent with our measurement from the times of superhump maxima (Figure 12). A phase diagram of the data from the plateau phase, folded on $P_{sh}$ is shown in Figure 11.

Again, we prefer to adopt our value of $P_{sh}$ from the time of maxima analysis as $P_{sh}$ = 0.08343(21) d.

**Searching for the orbital period**

To our knowledge, no direct measurement of the orbital period of CP Dra has been published, but $P_{orb}$ = 0.0816 d has been estimated from the superhump period using an empirical relationship between $P_{sh}$ and $P_{orb}$ [14, 23]. Although we could find no evidence for an orbital hump superimposed on the underlying superhumps by visual inspection of the light curves of the three superoutbursts, careful period analysis has in the past revealed subtle orbital signals in other SU UMa systems during superoutburst.

In the case of the 2009 superoutburst, we took the Lomb-Scargle power spectrum of the data around the superhump signal, between 0.07 and 0.09 d (top panel of Figure 14), where the largest peak corresponds to the superhump signal and pre-whitened with $P_{sh}$ to give the power spectrum in the bottom panel in Figure 14. This still has a prominent peak near the value of $P_{sh}$, as well as its 1 c/d alias, probably due to incomplete removal of the superhump signal because of its change of frequency during the outburst. The next two prominent signals are observed at P = 0.08084(86) d and 0.08803(79) d and which are 1 c/d aliases. We suggest that the signal at P = 0.08084(86) d, or ~1.94 h, is the orbital period (the other signal is well above the measured value of $P_{sh}$ and is therefore most unlikely to be the orbital period).

Taking the mean superhump period of the 2009 outburst, $P_{sh}$ = 0.08343(21), and the proposed orbital period, $P_{orb}$ = 0.08084(86) d, we calculate the superhump period excess ε = 0.032(3). Such value is consistent with other SU UMa systems of similar orbital period, although we note that it is at the lower end of the range [1]. For example KP Cas has a similar $P_{orb}$ = 0.0814(4) d to CP Dra, but its period excess is higher at ε = 0.048(5) [24], which is a more typical ε value. On the other hand, BR Lup has $P_{orb}$ = 0.0795(2) d and ε = 0.0340(40) [25], which is much closer to the value we found for CP Dra. If we assume that CP Dra has a white dwarf of ~0.75 solar masses, as is typical for SU UMa systems, then we can estimate the secondary to primary mass ratio, q, from the empirical relationship ε = 0.18*q + 0.29*q2 [25] as q = 0.14(2).

We also searched the data from 2001 and 2002 for signs of an orbital signal, but in these cases removal of the superhump signal left only very weak signals, none of which has any significant relationship to the possible orbital period found in the 2009 outburst.

**Discussion**

The similarity of the 2001 and 2009 superoutbursts is remarkable in terms of the profile of the outburst light curve as well as the evolution of the superhump amplitude and period. To investigate this further, we aligned the rapid decline phases of both outbursts, which are the most obvious and discrete features of the profiles, by adding JD 2949.021 to all times for the 2001 outburst and plotted them in the top panel of Figure 15. This shows that the two outburst profiles are virtually coincidental; it also exemplifies the fact that the 2001 outburst was caught during an earlier stage in the outburst. Furthermore in both outbursts the superhump period decrease was broadly similar: -9.2(9) x $10^{-4}$ in 2001 and -4.1(9) x $10^{-4}$ in 2009. A similar decrease of -2.2 x $10^{-4}$ was also reported for the 2003 superoutburst [14] and -1.35(95) x $10^{-3}$ in 2002. We also superposed the O-C data from 2001 and 2009 (bottom panel of Figure 15) and again note the very similar trend between the two outbursts. We analysed the superhump period change for the combined O-C data from 2001 and 2009, omitting the first three points from the 2001 outburst when the period was changing rapidly, and found a $P_{sh}$ decrease of -5.0(7) x $10^{-5}$ during the outburst. An alternative interpretation is that there are two discrete superhump regimes operating during the outbursts, rather than a continuous change, in that there appears to be a discontinuity in the O-C diagram (Figure 15, bottom panel) at around JD 2454916 to 2454917 when the superhump period abruptly evolves to a slightly shorter period which remains constant during the subsequent stages of the outburst. This discontinuity appears to coincide to the point in the outburst light curve (Figure 15, top panel) when a slow fading trend begins to set in during the plateau phase. A similar discontinuity has been observed in several other, but by no means all, long $P_{sh}$ systems [14]. Examples of such a system is SDSS J162718.39+120435.0 ($P_{sh}$ = 0.10993(7) d) [26] and V452 Cas ($P_{sh}$ = 0.08870(2) d), where the discontinuity also occurred at the onset of the slow decline from the plateau phase [18].

Although the $P_{sh}$ evolves during outbursts of CP Dra, as discussed above, we note the similarity of the mean $P_{sh}$ during the 4 characterised superoutbursts:

2001  $P_{sh}$ = 0.08335(31) d
2002  $P_{sh}$ = 0.08348(32) d
2003  $P_{sh}$ = 0.08348(10) d [14]
2009  $P_{sh}$ = 0.08343(21) d

Of these 4 superoutbursts, only the one in 2001 was observed in its early stages, during the rise to superoutburst. Here we noted that the superhump period was significantly longer (~6%) than during the rest of the superoutburst. The appearance of long-period superhumps during the rise to the superoutburst maximum is noteworthy, as this phase is more often characterised by the appearance of orbital humps [2]. One SU UMa system in which they were observed is HS 0417+7445 (= 1RXS J042332+745300), where $P_{sh}$ was ~9% longer during the rise [27]. We note that the $P_{orb}$ = 0.07531(8) d of HS 0417+7445 [27] is similar to that of CP Dra, but whether this behaviour is actually linked to $P_{orb}$ or the similar values of $P_{orb}$ is coincidental is not known since no global study of superhump evolution during the rise to superoutburst has been carried out, although Kato *et al.* note that a longer superhump period exists during the early evolutionary stage of some superoutbursts [14]. Unfortunately it is rather rare for superoutbursts to be caught during the rise phase, due to the brevity of such phases.

**Conclusions**

Analysis of observations of CP Dra between February 2001 and April 2009 has revealed 15 outbursts, of which at least eight were superoutbursts. The supercycle length is 230+/- 56 d.

We report photometry of the 2001, 2002 and 2009 superoutbursts. Those in 2001 and 2009 were the best observed and they were remarkably similar to each other in terms of the profile of the outburst light curve and the evolution of the superhumps. The outburst amplitude was 5.5 magnitudes and the $P_{sh}$ during the plateau phase was measured at 0.08335(31) and 0.08343(21) d, respectively. In both cases $P_{sh}$ decreased during the course of the outburst; whilst analysis of the individual outbursts was consistent with a continuous period reduction, there is some evidence from the combined data from the two outbursts that there was an abrupt change corresponding to the point in the plateau where a slow fade sets in. Superhumps of 0.3 magnitude peak-to-peak amplitude were present at the height of each outburst and these gradually diminished during the outburst. Rather less coverage of the 2002 superoutburst was obtained, but we measured the $P_{sh}$ during the plateau as = 0.08348(32) d, which is also consistent with the other two superoutbursts.

The 2001 superoutburst was caught during the rise to maximum and during this period we found that the $P_{sh}$ was significantly longer than during the subsequent stages of the outburst.

Analysis of time resolved photometry from the 2009 outburst has revealed evidence for an orbital period of 0.08084(86) d, giving a fractional superhump period excess of 0.032(3) d.

## Acknowledgements

The authors gratefully acknowledge the use of observations from the AAVSO International Database, from the database of the Center for Backyard Astrophysics (CBA) and from the BAA Recurrent Objects Programme (ROP) contributed by observers worldwide. We are grateful to Professor Joe Patterson (Columbia University, USA) for allowing us to use the CBA data as well as for orchestrating the 2001 campaign and Gary Poyner for furnishing data from the ROP. We thank Wolfgang Renz for locating a copy of the paper from Peremennye Zvezdy Prilozhenie and Dennis Kruchinin for translating it from the Russian. Dan Green (Director CBAT) helpfully scanned IAUC 2381 and 2383 for us and made them available online. At our request, Jean Marie Llapasset (Perpignan, France) kindly carried out photometry on his images of SN2008fv in NGC 3147 which showed CP Dra in outburst and which was important in identifying a hitherto unreported outburst in 2008.

## Addresses:

JS: "Pemberton", School Lane, Bunbury, Tarporley, Cheshire, CW6 9NR, UK [bunburyobservatory@hotmail.com]
DBo: 5 Silver Lane, West Challow, Wantage, Oxon, OX12 9TX, UK [drsboyd@dsl.pipex.com]
DBu, Conder Brow Observatory, Littlefell Lane, Lancaster, LA2 0RQ , UK [buczynski8166@btinternet.com]
BM, The King's University College; Center for Backyard Astrophysics (Alberta), Edmonton, Alberta, Canada T6B 2H3 [bemart1@shaw.ca]
DM, Center for Backyard Astrophysics (Norwich), Norwich, USA [dpmessier@aol.com]
IM: Furzehill House, Ilston, Swansea, SA2 7LE, UK [furzehillobservatory@hotmail.com]
AO, Verkkoniementie 30, FI-40950 Muurame, Finland [arto.oksanen@jklsirius.fi]
JP: Nollenweg 6, 65510 Idstein, Germany [j.pietz@arcor.de]
DS, Center for Backyard Astrophysics (East), 6-G Ridge Road Greenbelt, MD20770, USA [dskillman@comcast.net]
BS: Alan Guth Observatory, Koningshofbaan 51, Hofstade, Aalst, Belgium [staels.bart.bvba@pandora.be]
TV, Center for Backyard Astrophysics (Belgium), Walhostraat 1A, B-3401 Landen, Belgium [tonny.vanmunster@cbabelgium.com]

| JD | Calendar date |
|---|---|
| 2441441 | 1972 May 3 |
| 2441826 | 1973 May 23 |
| 2442122 | 1974 Mar 15 |
| 2442423 | 1975 Jan 10 |
| 2442901 | 1976 May 2 |
| 2443216 | 1977 Mar 13 |

**Table 1: Outbursts of CP Dra in the 1970s**
Data from reference 7

| Detection date | JD | Magnitude at maximum | Outburst duration (d) | Superhumps detected? | Comment |
|---|---|---|---|---|---|
| 2001 Feb 23 | 2451964.4 | 14.3 | >10 | Y | This paper |
| 2002 Mar 27 | 2452361.3 | 15.3 | >7 | Y | This paper |
| 2003 Jan 8 | 2452648.4 | 14.5 | >4 | Y [14] | |
| 2003 Mar 21 | 2452720.3 | 15.6 | ? | | Only 2 observations |
| 2003 Nov 23 | 2452965.0 | 14.6 | >7 | | Nogami comments this was a superoutburst [15] |
| 2004 Apr 12 | 2453108.3 | 15.6 | ? | | Single observation |
| 2005 Feb 28 | 2453410.3 | 14.5 | >10 | | |
| 2005 Aug 14 | 2453597.4 | 15.1 | ? | N [16] | |
| 2006 Jan 1 | 2453737.4 | 14.7 | ? | N [17] | |
| 2006 Jun 7 | 2453894.4 | 14.3 | >8 | | |
| 2007 Jul 14 | 2454296.4 | 14.8 | ? | | Single observation |
| 2008 Feb 10 | 2454507.3 | 15.1 | >13 | | |
| 2008 Aug 22 | 2454701.4 | 15.3 | ? | | Single observation |
| 2008 Oct 27 | 2454766.6 | 14.6 | ? | | Single observation |
| 2009 Mar 21 | 2454912.4 | 14.6 | ~14 | Y | This paper |

**Table 2: Outbursts of CP Dra between February 2000 and April 2009**
Confirmed and probable superoutbursts are highlighted in grey

| Detection date | JD | ΔT (d) |
|---|---|---|
| 2001 Feb 23 | 2451964.6 | |
| 2002 Mar 27 | 2452361.3 | 396.7 |
| 2003 Jan 8 | 2452648.4 | 287.1 |
| 2003 Nov 23 | 2452965.0 | 316.6 |
| 2005 Feb 28 | 2453410.3 | 445.3 |
| 2006 Jun 7 | 2453894.4 | 484.1 |
| 2008 Feb 10 | 2454507.3 | 612.9 |
| 2009 Mar 21 | 2454912.4 | 405.1 |

**Table 3: Superoutbursts of CP Dra**

| Observer | Telescope | CCD (unfiltered) |
|---|---|---|
| Vanmunster | 0.35 m SCT | SBIG ST-7XME |
| Buczynski | 0.33 m reflector | Starlight Xpress SXL8 |
| Martin | 0.32 m reflector | SBIG ST-7E |
| Skillman | 0.66 reflector | Apogee AP-7 |
| Messier | 0.25 m SCT | SBIG ST-8 |
| Pietz | 0.20 m SCT | SBIG ST-6B |
| Oksanen | 0.40 m Ritchey-Chretien | SBIG STL-1001E |
| Shears | 0.28 m SCT | Starlight Xpress SXVF-H9 |
| Miller | 0.35 m SCT | Starlight Xpress SXVF-H16 |
| Boyd | 0.35 m SCT | Starlight Xpress SXV-H9 |
| Staels | 0.28 m SCT | Starlight Xpress MX716 |

**Table 4: Equipment used**

| Start time | Duration (h) | Observer |
|---|---|---|
| 2451965.296 | 7.54 | Vanmunster |
| 2451965.375 | 4.80 | Buczynski |
| 2451966.634 | 2.33 | Martin |
| 2451967.310 | 4.58 | Vanmunster |
| 2451967.611 | 9.21 | Skillman |
| 2451967.654 | 7.13 | Messier |
| 2451969.667 | 6.00 | Skillman |
| 2451971.611 | 7.90 | Martin |
| 2451972.608 | 10.58 | Martin |
| 2451974.283 | 5.47 | Vanmunster |

**Table 5: Log of time-series observations, 2001 outburst**

| Superhump cycle no. | Time of maximum HJD | Error (d) |
|---|---|---|
| -16 | 2451965.3473 | 0.0003 |
| -15 | 2451965.4355 | 0.0007 |
| -14 | 2451965.5242 | 0.0005 |
| 0 | 2451966.7147 | 0.0023 |
| 8 | 2451967.3846 | 0.0005 |
| 9 | 2451967.4663 | 0.0011 |
| 10 | 2451967.5505 | 0.0005 |
| 11 | 2451967.6374 | 0.0009 |
| 12 | 2451967.7197 | 0.0009 |
| 12 | 2451967.7258 | 0.0008 |
| 13 | 2451967.8015 | 0.0033 |
| 13 | 2451967.8116 | 0.0008 |
| 13 | 2451967.8013 | 0.0035 |
| 14 | 2451967.8878 | 0.0027 |
| 14 | 2451967.8965 | 0.0005 |
| 14 | 2451967.8855 | 0.0004 |
| 15 | 2451967.9721 | 0.0032 |
| 38 | 2451969.8866 | 0.0008 |
| 59 | 2451971.6400 | 0.0012 |
| 60 | 2451971.7212 | 0.0027 |
| 61 | 2451971.8050 | 0.0014 |
| 62 | 2451971.8914 | 0.0034 |
| 71 | 2451972.6398 | 0.0039 |
| 72 | 2451972.7187 | 0.0051 |
| 73 | 2451972.8025 | 0.0052 |
| 74 | 2451972.8864 | 0.0057 |
| 75 | 2451972.9715 | 0.0057 |

**Table 6: Times of superhump maximum in 2001**

| Start time | Duration (h) | Observer |
|---|---|---|
| 2452361.324 | 3.94 | Pietz |
| 2452362.320 | 3.98 | Pietz |
| 2452363.360 | 3.70 | Pietz |
| 2452365.435 | 1.63 | Pietz |

**Table 7: Log of time-series observations, 2002 outburst**

| Superhump cycle no. | Time of maximum HJD | Error (d) |
|---|---|---|
| 0 | 2452361.3729 | 0.0006 |
| 1 | 2452361.4556 | 0.0007 |
| 12 | 2452362.3766 | 0.0008 |
| 13 | 2452362.4594 | 0.0009 |
| 24 | 2452363.3786 | 0.0011 |
| 25 | 2452363.4620 | 0.0012 |
| 49 | 2452365.4625 | 0.0015 |

**Table 8: Times of superhump maximum in 2002**

| Start time | Duration (h) | Observer |
|---|---|---|
| 2454915.421 | 4.92 | Oksanen |
| 2454917.322 | 7.13 | Shears |
| 2454917.393 | 5.33 | Miller |
| 2454917.398 | 3.94 | Boyd |
| 2454918.387 | 1.13 | Miller |
| 2454919.385 | 4.27 | Boyd |
| 2454919.388 | 1.27 | Shears |
| 2454919.483 | 5.04 | Miller |
| 2454920.308 | 6.94 | Staels |
| 2454920.341 | 3.29 | Boyd |
| 2454921.371 | 2.13 | Staels |
| 2454922.303 | 2.56 | Staels |
| 2454922.390 | 2.02 | Boyd |
| 2454923.297 | 1.82 | Staels |
| 2454923.331 | 4.27 | Boyd |
| 2454923.457 | 3.31 | Miller |

**Table 9: Log of time-series observations, 2009 outburst**

| Superhump cycle no. | Time of maximum HJD | Error (d) |
|---|---|---|
| 0 | 2454915.4382 | 0.0004 |
| 1 | 2454915.5218 | 0.0002 |
| 2 | 2454915.6068 | 0.0004 |
| 24 | 2454917.4488 | 0.0006 |
| 24 | 2454917.4487 | 0.0014 |
| 25 | 2454917.5303 | 0.0003 |
| 25 | 2454917.5279 | 0.0011 |
| 48 | 2454919.4493 | 0.0016 |
| 49 | 2454919.5335 | 0.0006 |
| 49 | 2454919.5309 | 0.0004 |
| 50 | 2454919.6155 | 0.0003 |
| 59 | 2454920.3619 | 0.0020 |
| 59 | 2454920.3635 | 0.0007 |
| 60 | 2454920.4451 | 0.0015 |
| 60 | 2454920.4475 | 0.0003 |
| 61 | 2454920.5331 | 0.0006 |
| 84 | 2454922.4472 | 0.0006 |
| 95 | 2454923.3654 | 0.0012 |
| 96 | 2454923.4531 | 0.0004 |

**Table 10: Times of superhump maximum in 2009**

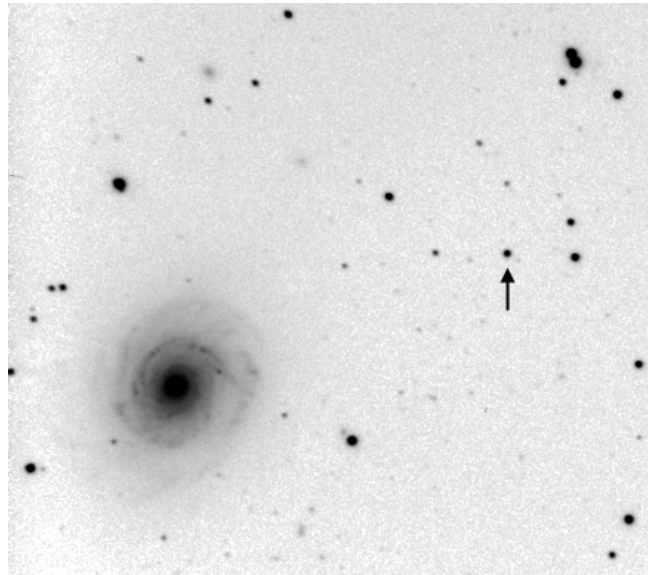

**Figure 1: CP Dra in outburst on 2009 March 29.90, with the nearby galaxy NGC 3147**
*(David Boyd)*

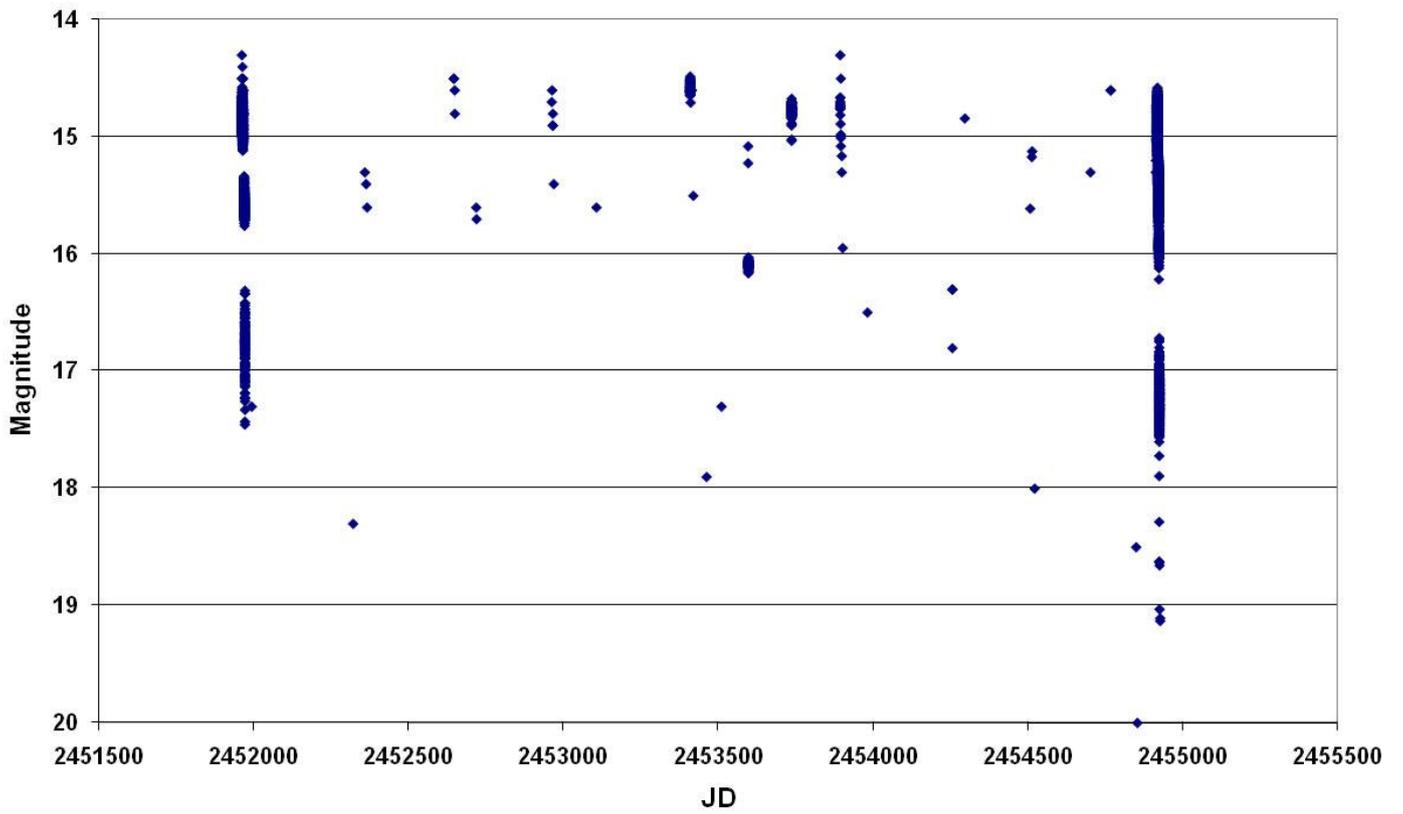

**Fig 2: Light curve of CP Dra between 2001 Feb and 2009 April**

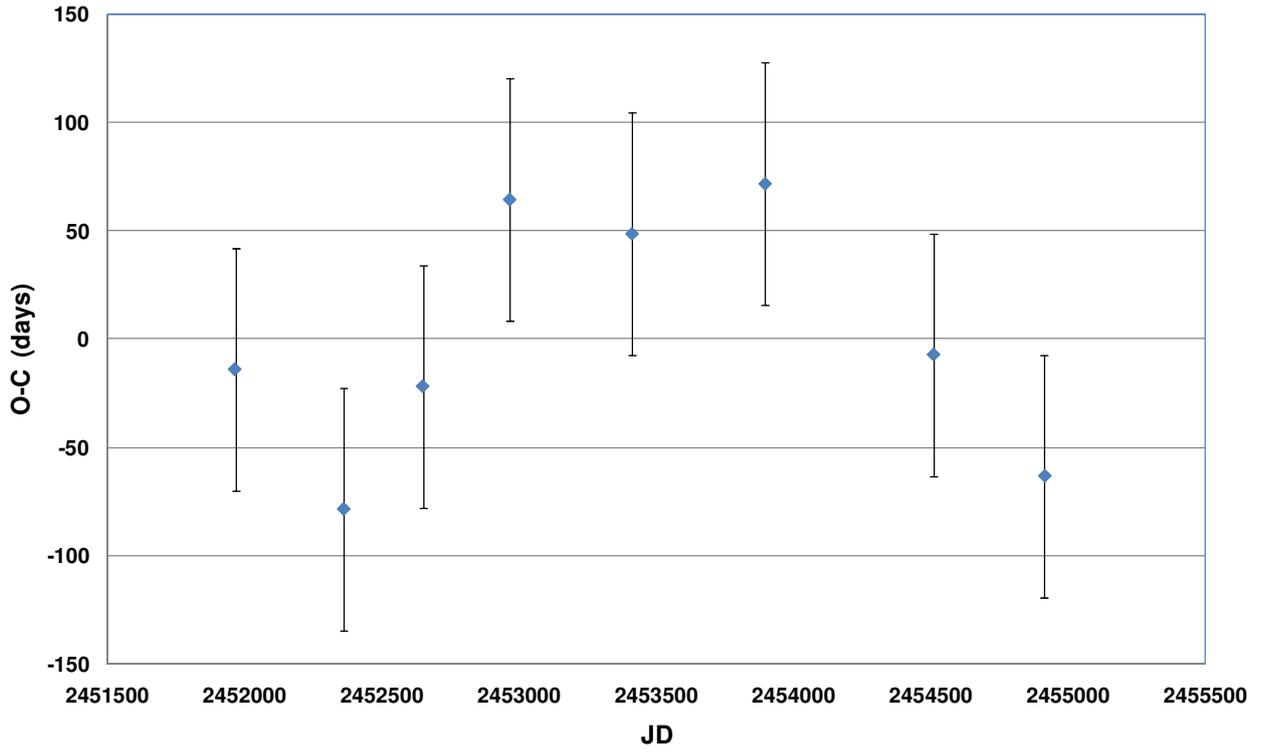

**Figure 3: O-C diagram of superoutbursts of CP Dra**
Error bars represent the standard deviation of the O-C residuals (56 d)

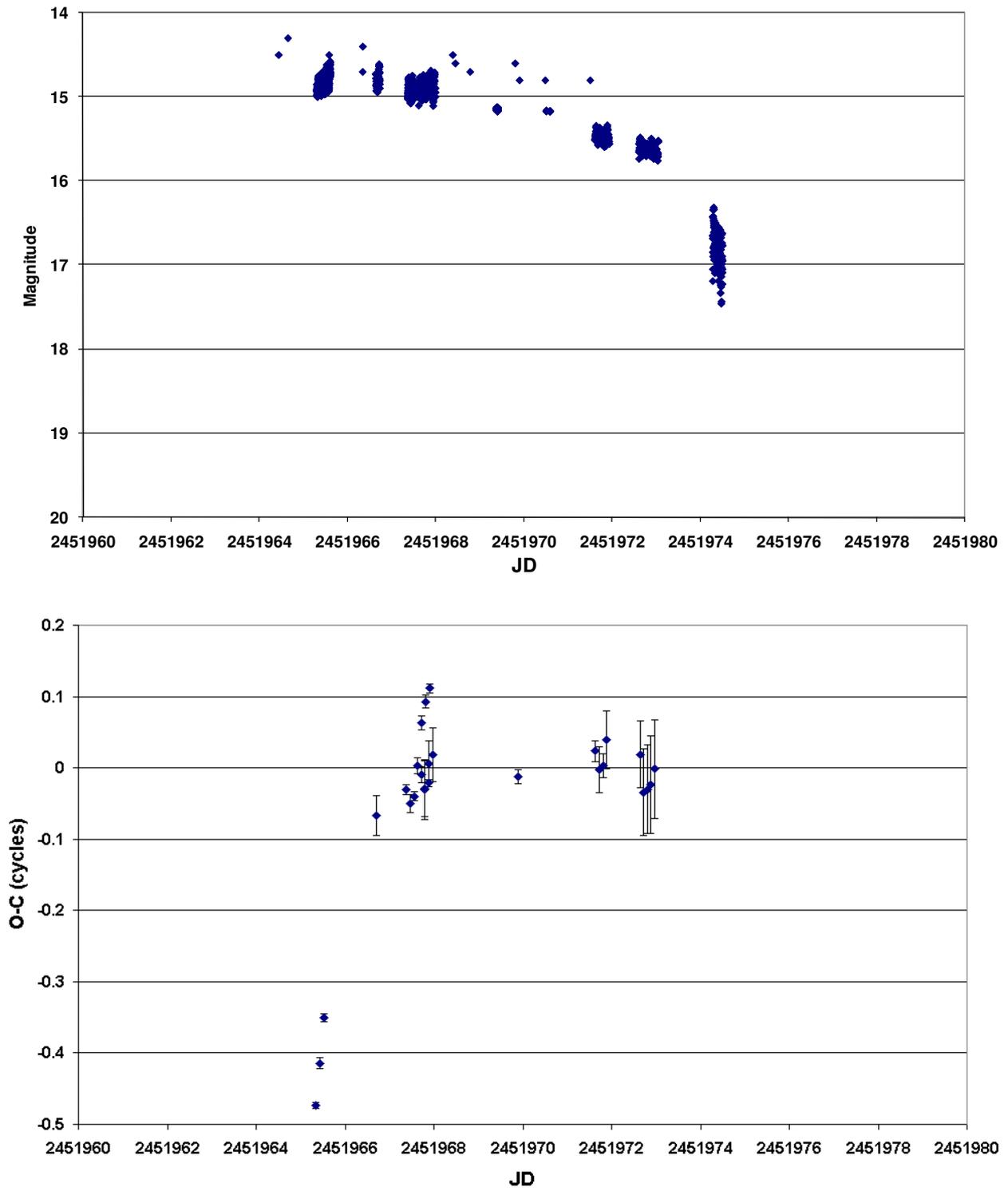

**Figure 4: Light curve of the 2001 outburst (top) and O-C diagram of superhump maxima (bottom)**

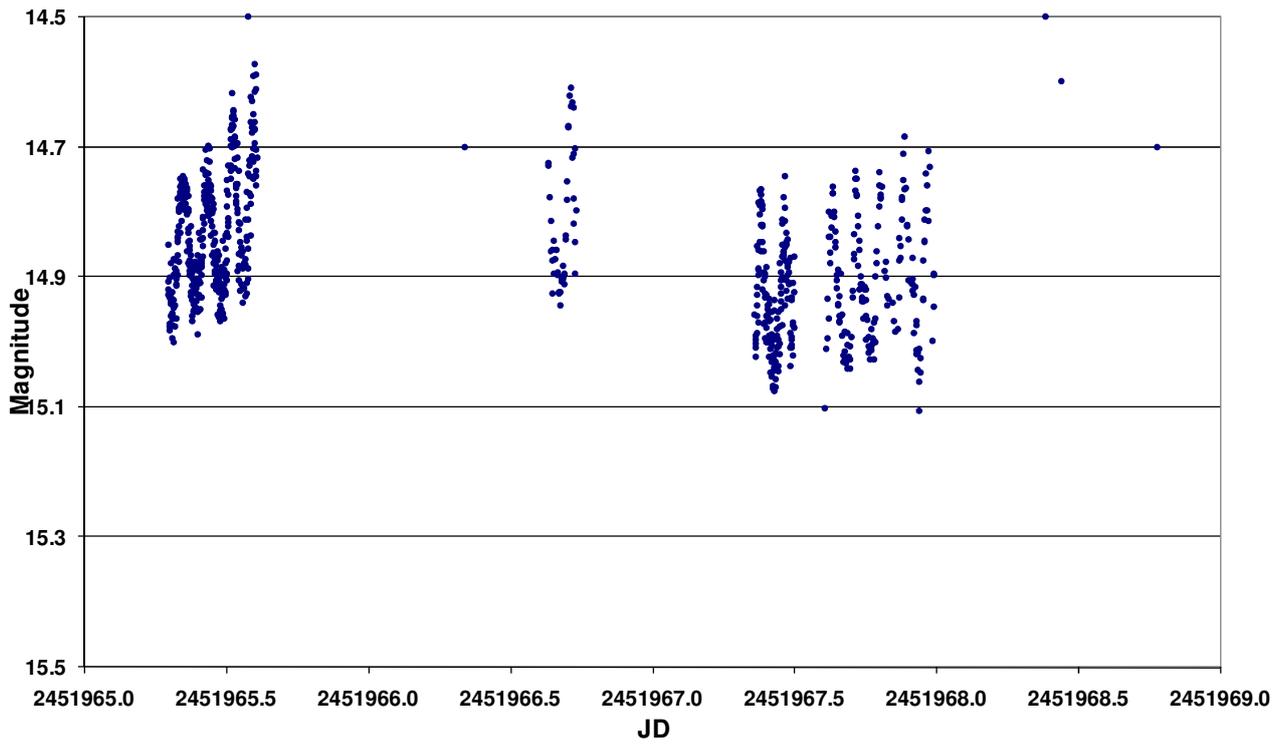

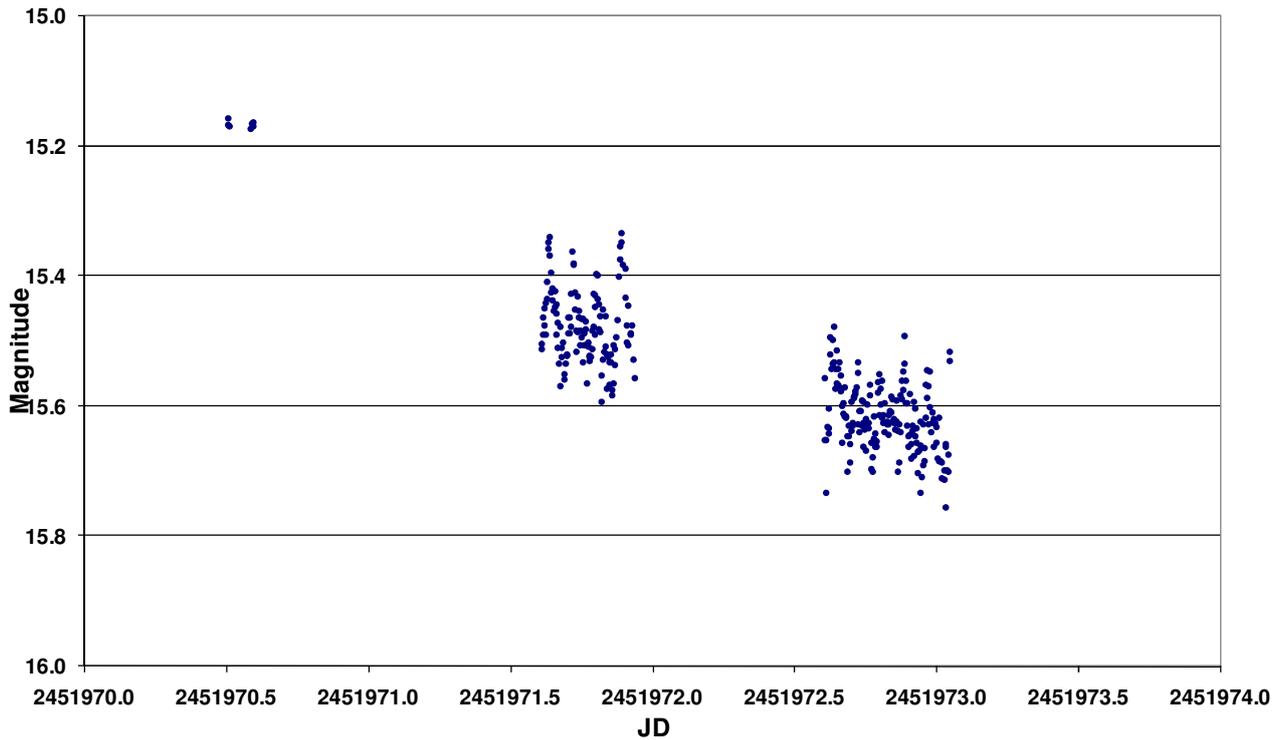

**Figure 5: Superhumps during the plateau phase of the 2001 outburst**

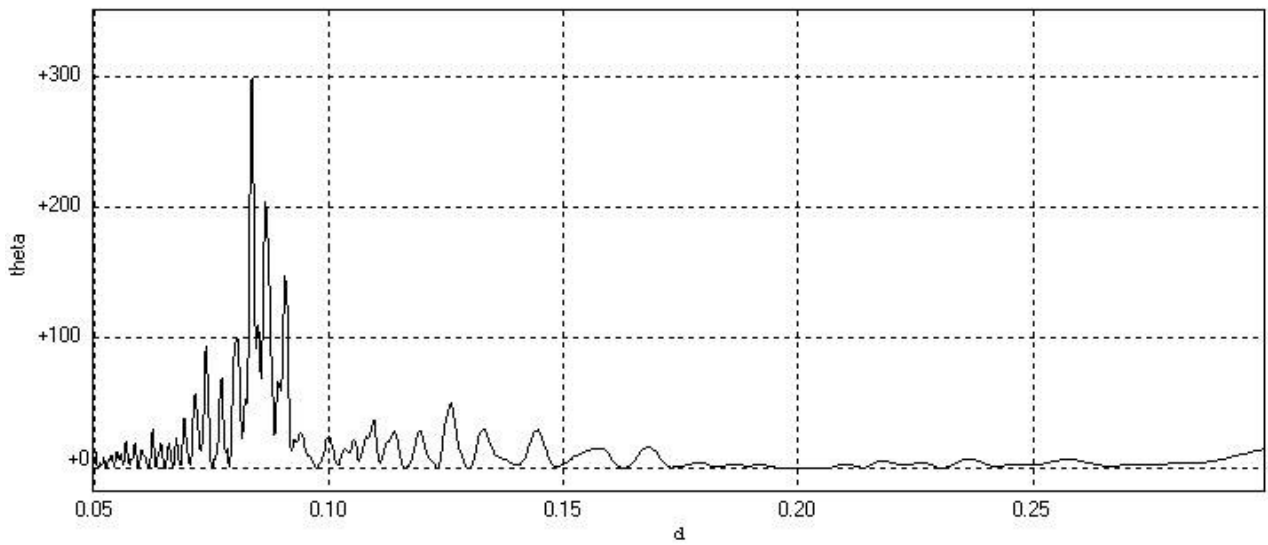

**Figure 6: Power spectrum of the 2001 outburst from Lomb-Scargle analysis**

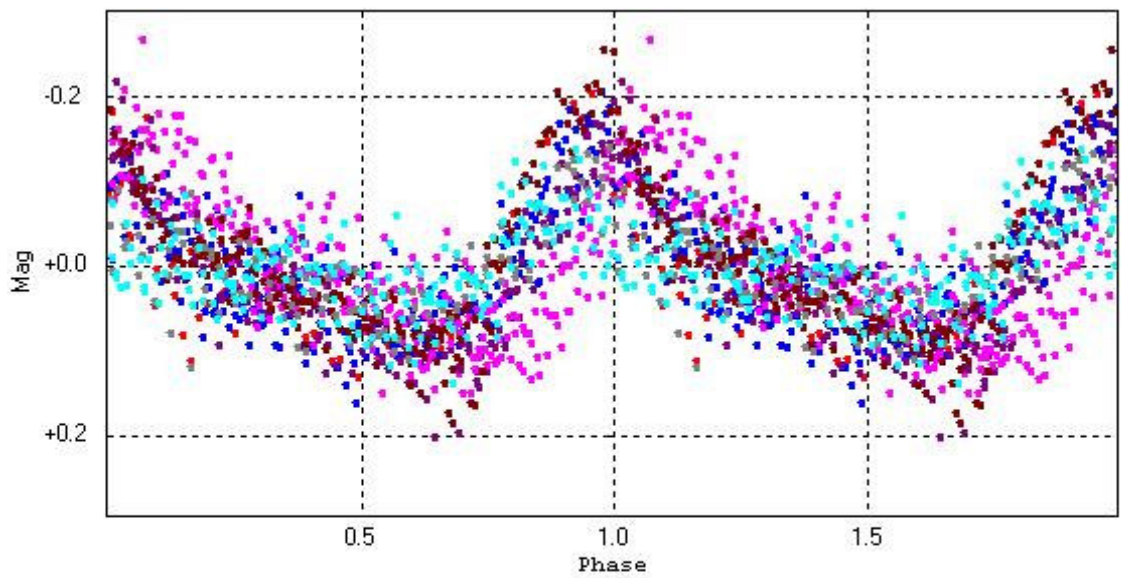

**Figure 7: Phase diagram of the 2001 outburst folded on $P_{sh}$ of 0.08345 d**

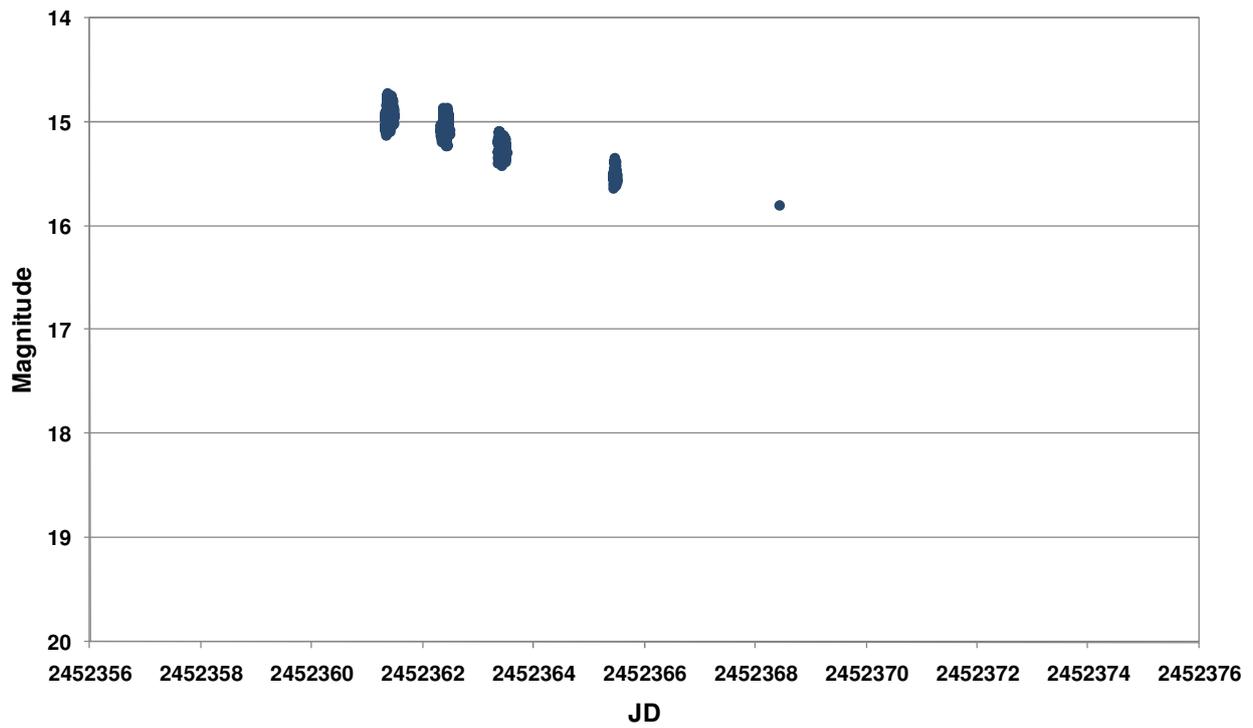

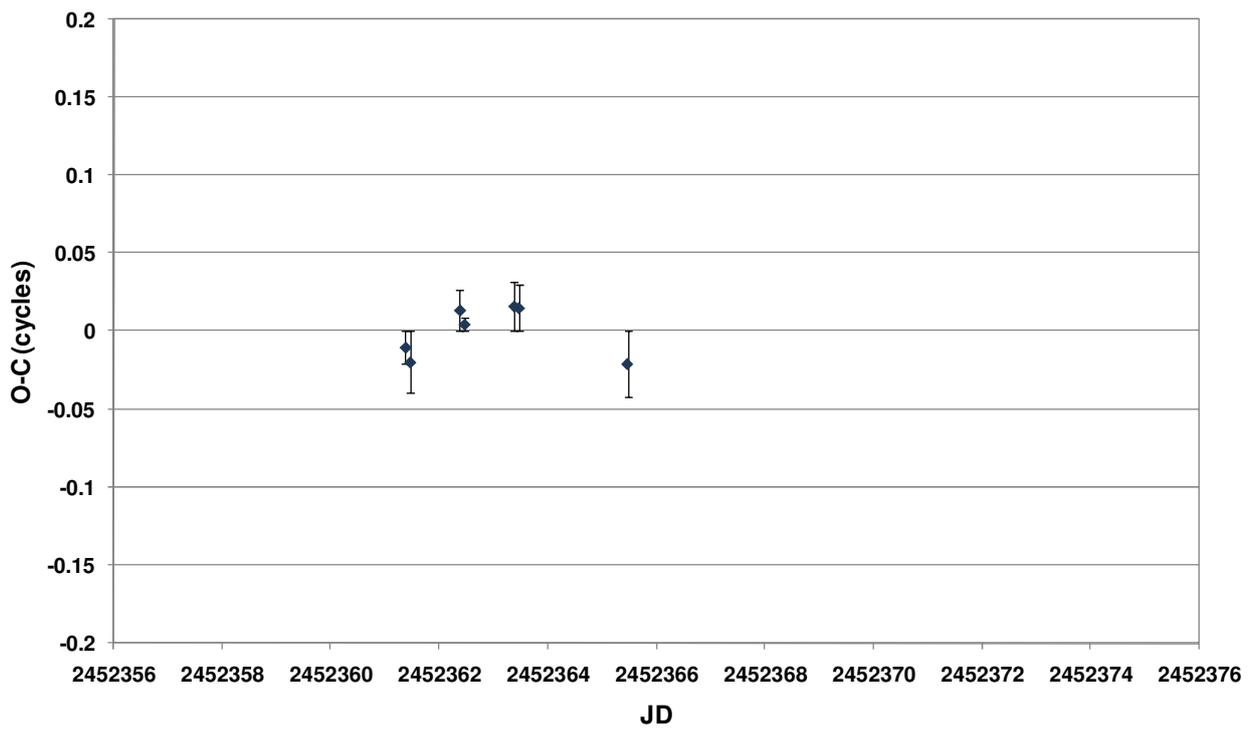

**Figure 8: Light curve of the 2002 outburst (top) and O-C diagram of superhump maxima (bottom)**

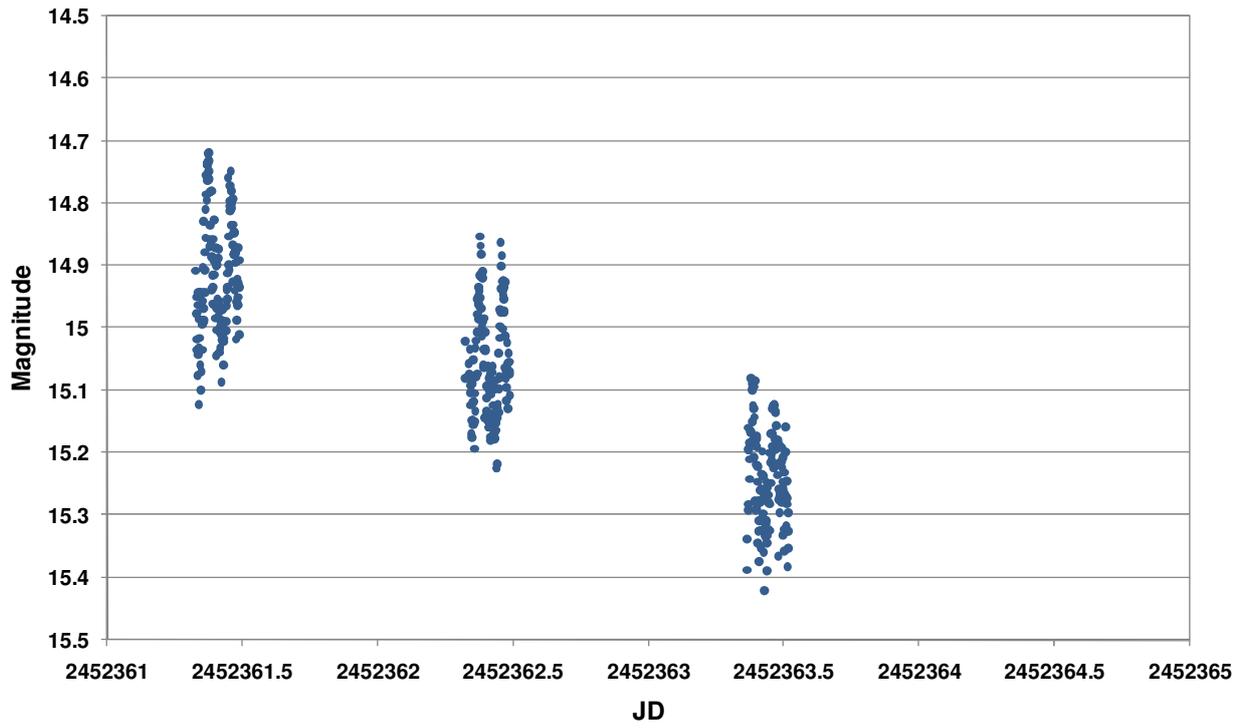

**Figure 9: Superhumps during the plateau phase of the 2002 outburst**

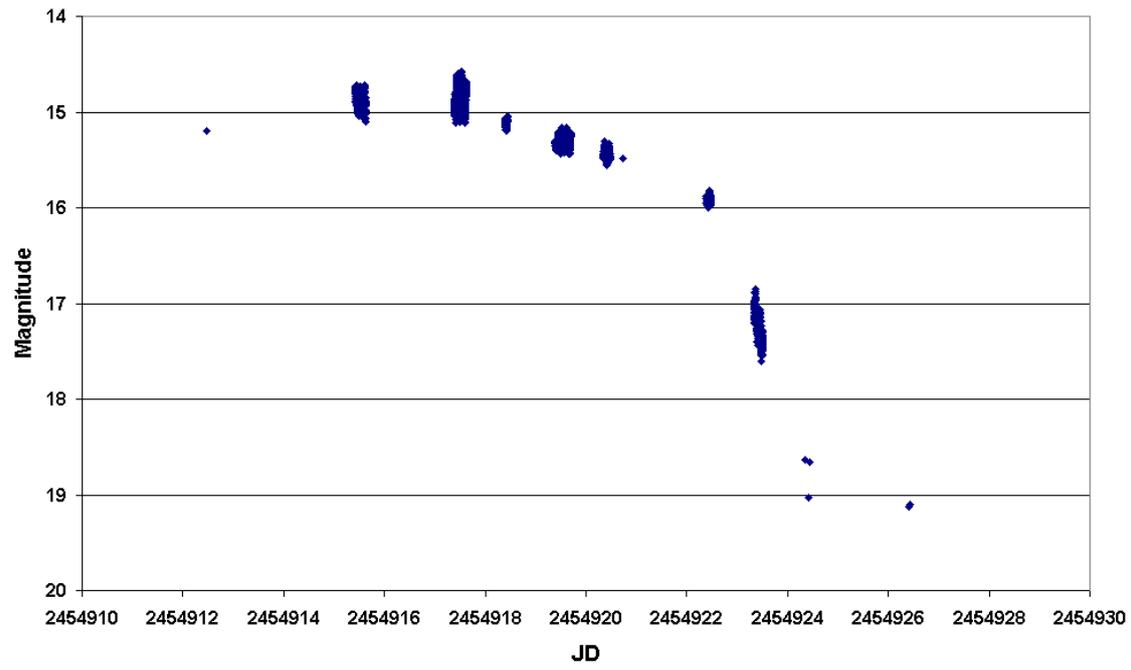

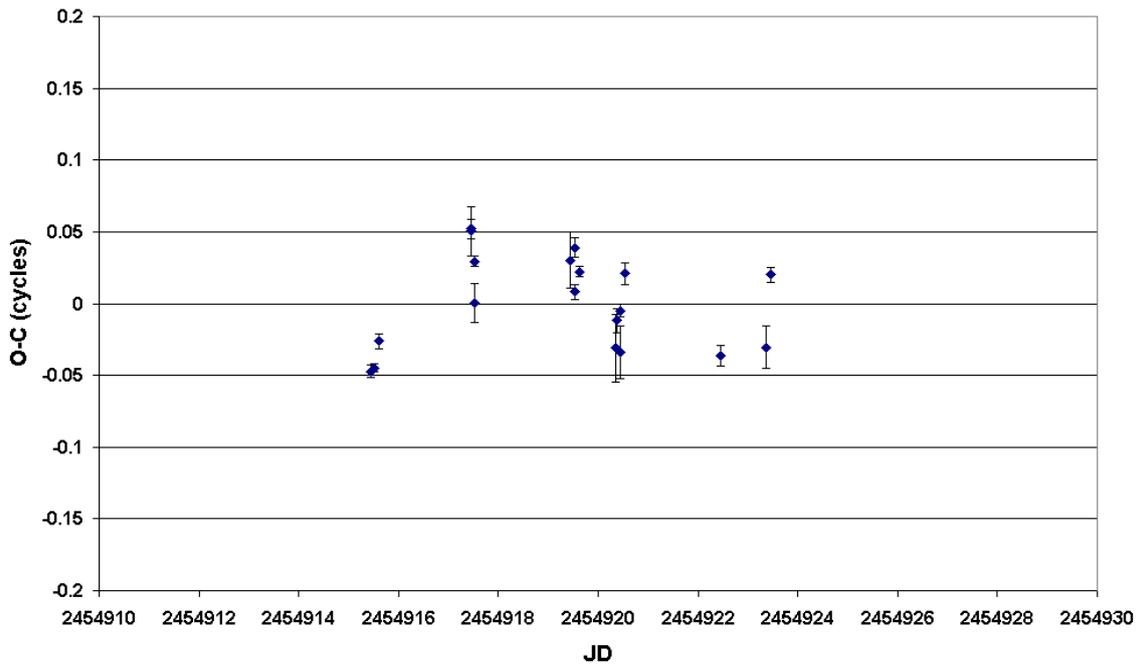

**Figure 10: Light curve of the 2009 outburst (top) and O-C diagram of superhump maxima (bottom)**

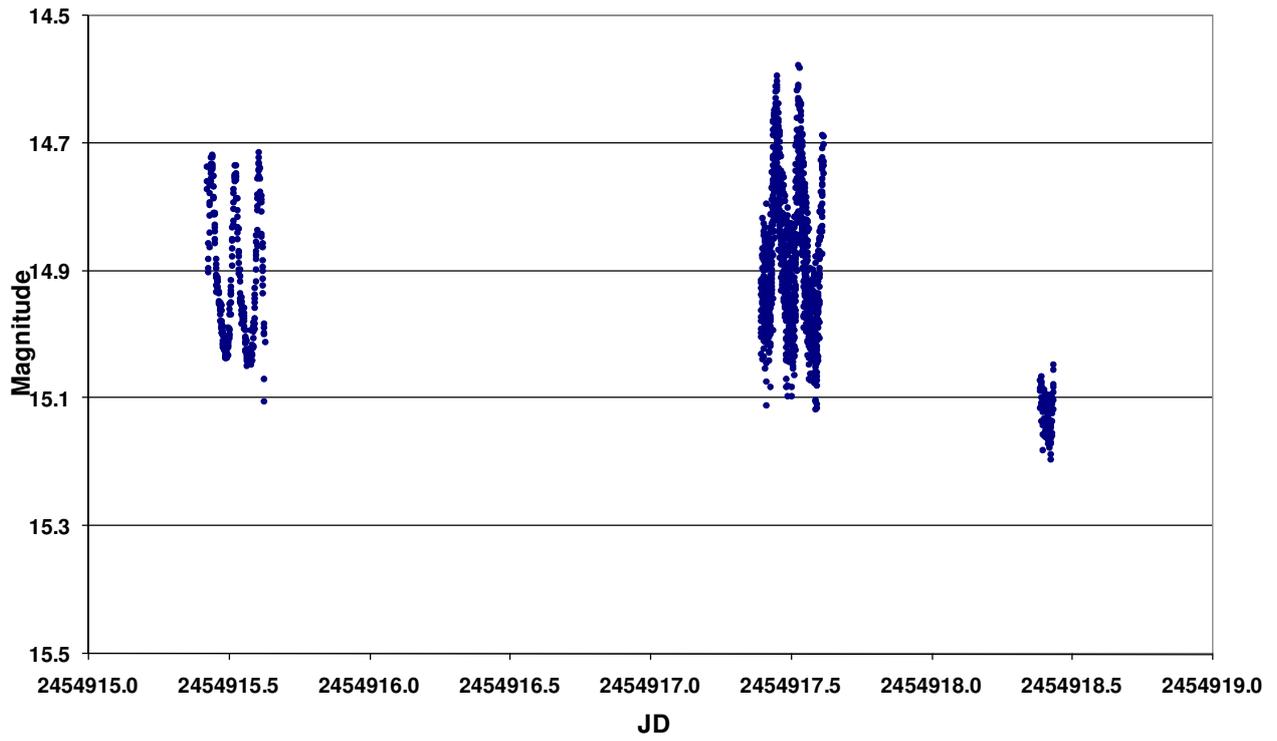

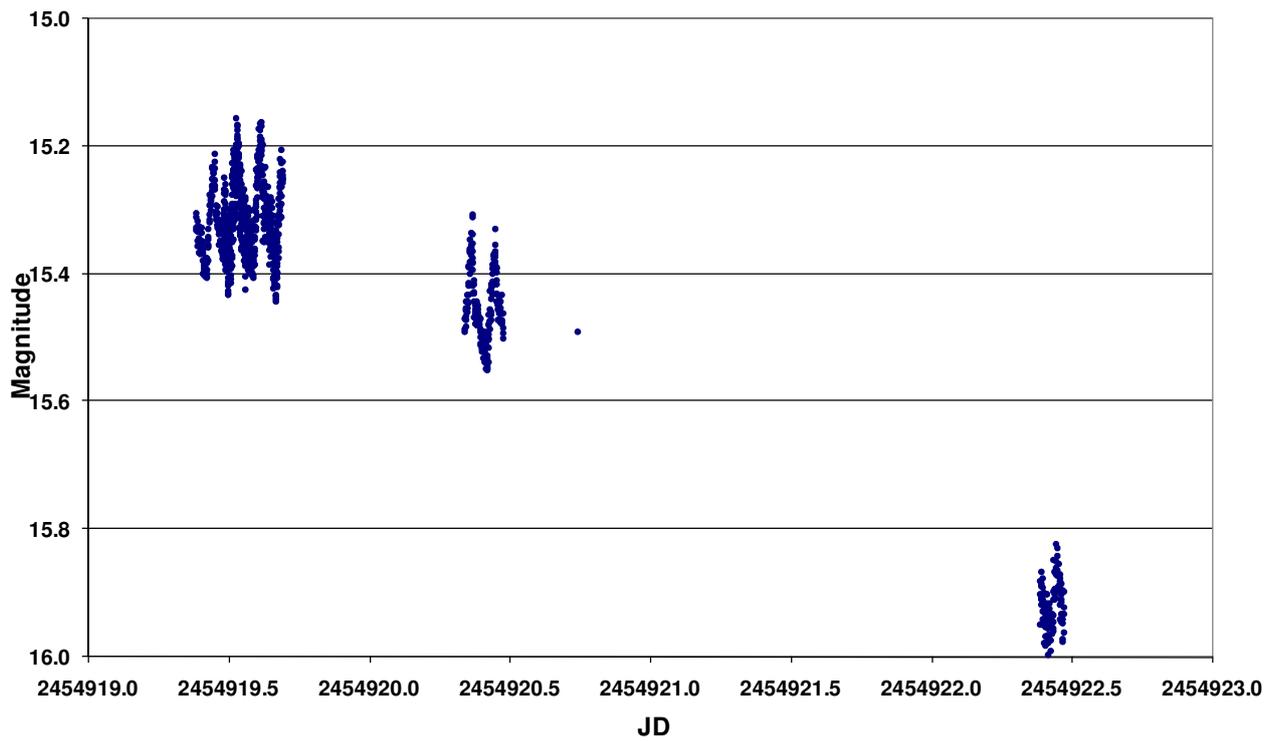

**Figure 11: Superhumps during the plateau phase of the 2009 outburst**

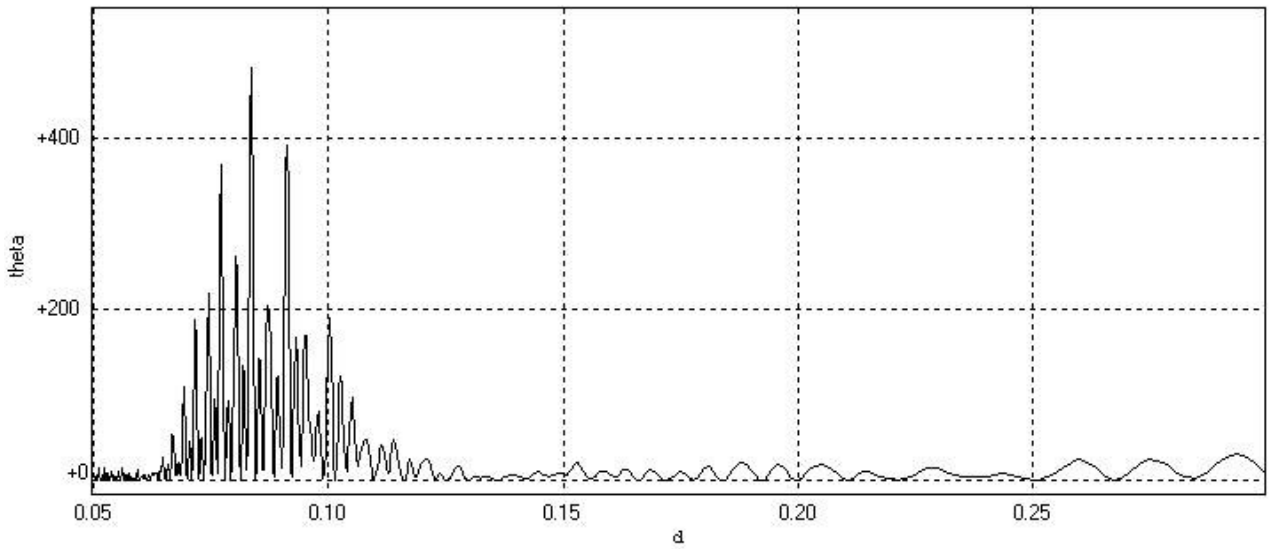

**Figure 12: Power spectrum of the 2009 outburst from Lomb-Scargle analysis**

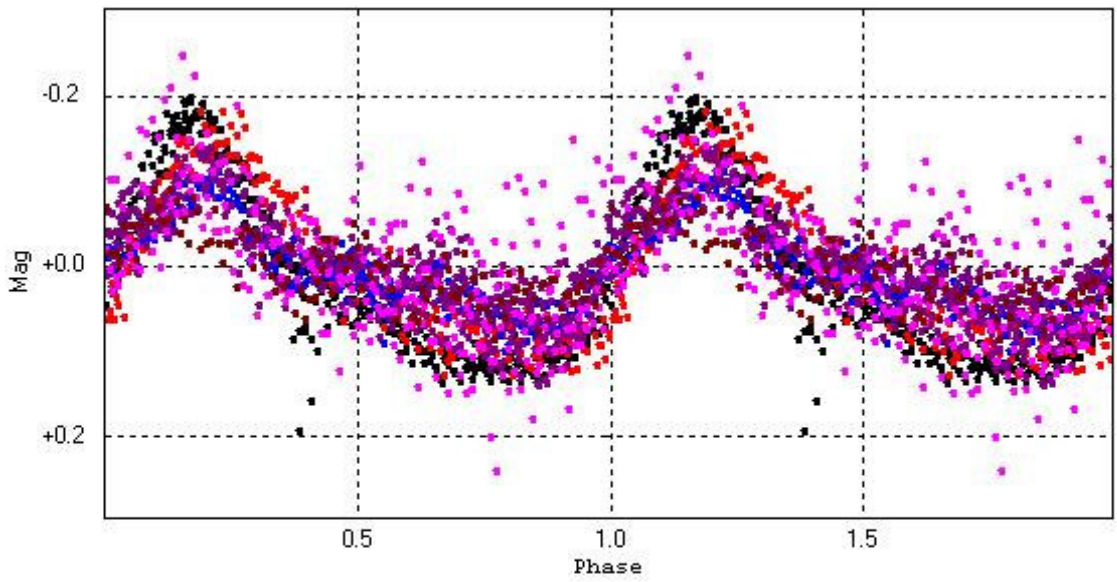

**Figure 13: Phase diagram of the 2009 outburst folded on $P_{sh}$ of 0.08348 d**

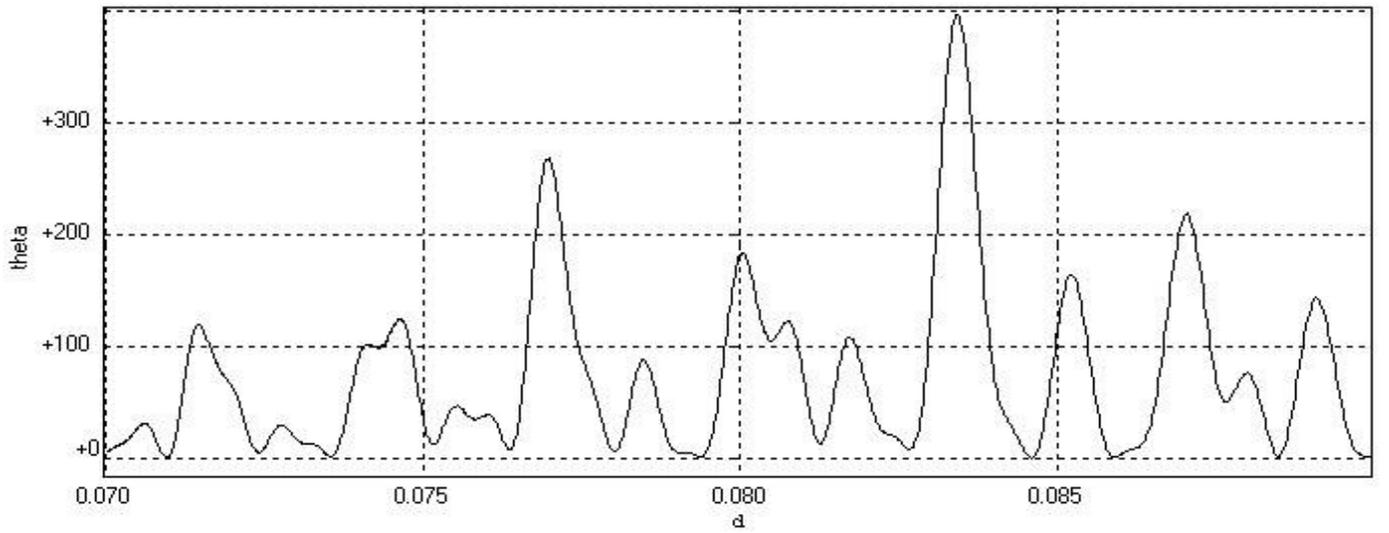

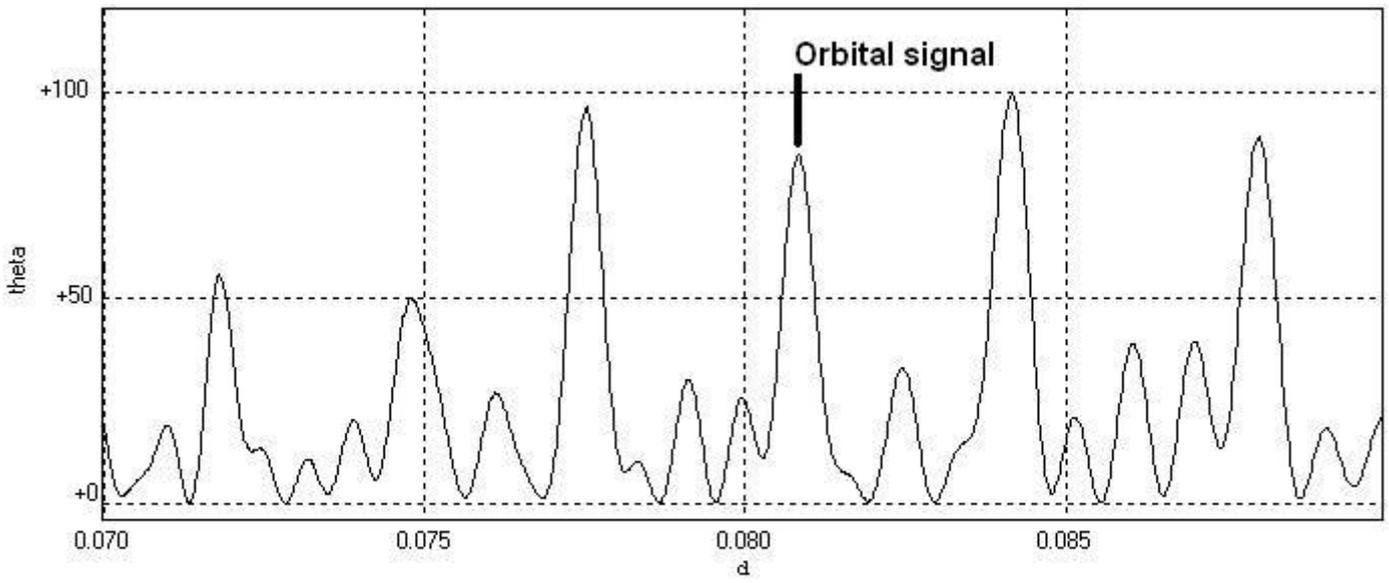

**Figure 14: Expanded power spectrum of the 2009 outburst from Lomb-Scargle analysis before (top) and after (bottom) removing the superhump signal at 0.08348 d**

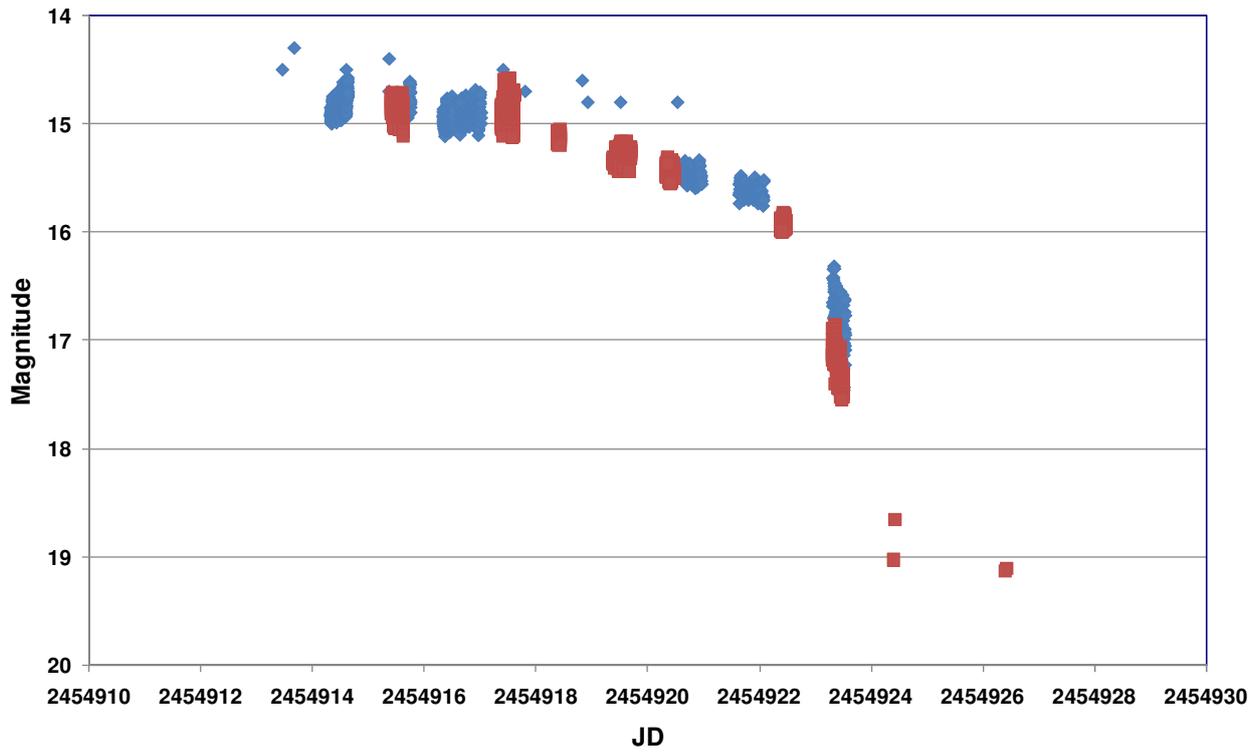

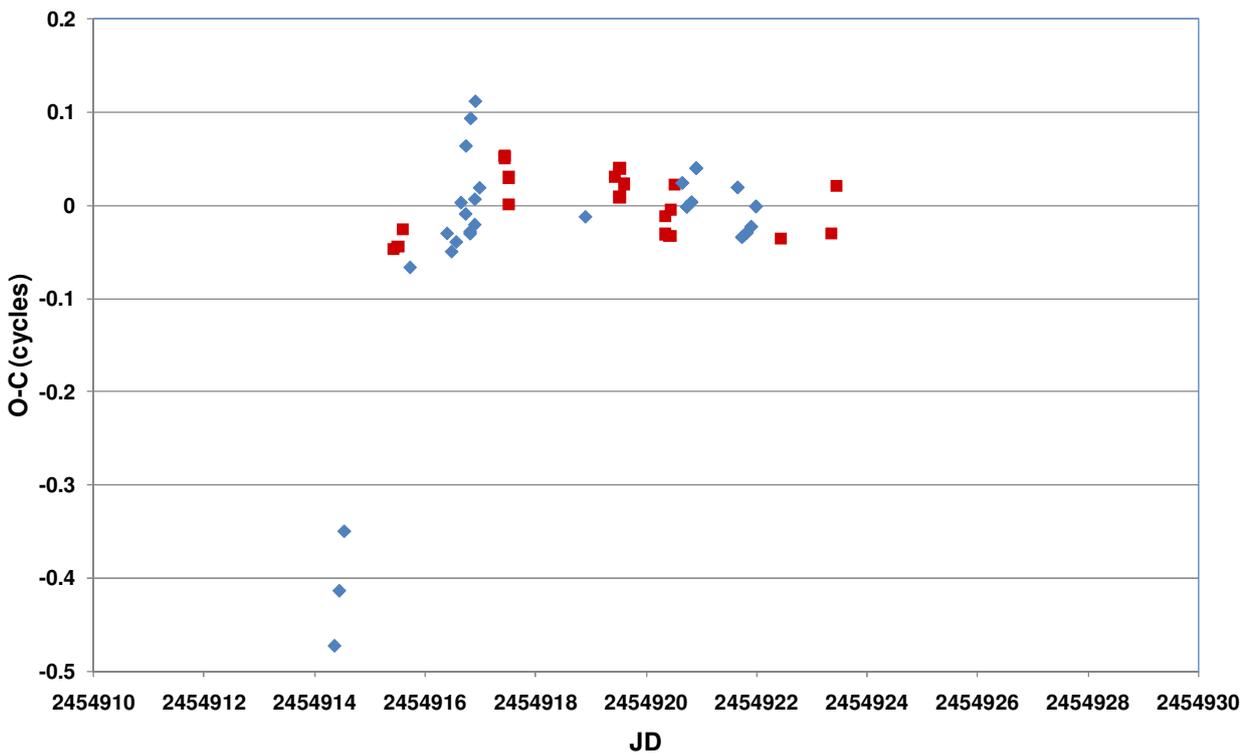

**Figure 15: Combined light curves (top) and O-C diagrams (bottom) of the 2001 and 2009 outbursts**

2001 data: blue, 2009 data: red. 2001 data were transformed by adding JD 2949.021